\documentclass[12pt]{article}
\usepackage{amsbsy}
\usepackage[dvips]{graphicx}
\title{Nuclear effects and  higher twists in $F_3$ structure function
}
\author{
S. A. Kulagin$^{1}$ and A. V. Sidorov$^{2}$ \\[0.5cm]
\small
${}^1$ Institute for Nuclear Research of the Russian Academy of Sciences,\\
\small
117312 Moscow, Russia
\\
\small
${}^2$ Bogoliubov Laboratory of Theoretical Physics,\\
\small
Joint Institute for Nuclear Research,  141980 Dubna, Russia
}                     
\begin{document}
%
\maketitle
\begin{abstract}
\noindent
We analyze the CCFR collaboration iron target data on the $xF_3$
structure function making particular emphasis on the extraction of the higher
twist contributions from data.
Corrections for nuclear effects are applied in order to extract data on the structure
function of the isoscalar nucleon.
Our analysis confirms the observation made earlier, that the higher twist terms depend
strongly on the level to which QCD perturbation theory analysis is applied.
We discuss the impact of nuclear effects on the higher twist term as well as
on the QCD scale parameter $\Lambda_{\overline{MS}}$ extracted from the fit
to data.\\[1cm]
\end{abstract} 
\newpage

\section{Introduction}
\label{intro}

In the present paper we report the results of our analysis of the CCFR
collaboration data \cite{CCFR} on the structure function $xF_3$.
The particular emphasis of the analysis is to constrain the higher twist
(HT) contributions to the structure function from data.
The HT effects in the $xF_3$
structure function are of
particular interest because of certain theoretical predictions
made in the framework of infrared renormalon
technique \cite{renormalon,h-function,renormrev,Zakharov}.
%
An attempt to constrain the HT terms from the CCFR/NuTeV
collaboration data was done in \cite{fit1,fit2}, where
the $F_3$ structure function was written as the sum of two terms,
\begin{eqnarray}
\label{LTplusHT}
xF_3(x,Q^2)=xF_3^{LT}(x,Q^2) +\frac{h(x)}{Q^2},
\end{eqnarray}
with $F_3^{LT}$ the leading twist contribution and $h/Q^2$ the HT term.
An important observation which follows from the analysis \cite{fit1,fit2,S1}
is that the magnitude of the HT term depends on the level to which the perturbation
theory analysis of $F_3^{LT}$ is applied.
If $F_3^{LT}$ is evaluated in a leading order (LO) renormalization group formalism
a large $h(x)$ appears from the fit. When a next-to-leading order (NLO) formalism
is used for $F_3^{LT}$ a somewhat smaller but still substantial contribution from
the HT term is needed. If $F_3^{LT}$ is evaluated to next-to-next-to-leading order
(NNLO) very little room is left for the HT term.

We note that QCD analysis of data implies that data are given
for isolated proton and neutron. In practice, due to the reason of
statistics, neutrino data are taken mainly on nuclear targets rather
than on isolated proton and neutron. For example,
the CCFR/NuTeV collaboration uses the iron target,
the IHEP-JINR Neutrino Detector uses the aluminum target \cite{IHEP-JINR},
and the forthcoming data from CHORUS
collaboration is obtained on the lead target \cite{oldeman}).
It is known from muon and electron DIS experiments, that nuclear effects are
quite essential in a wide kinematical region of $x$ and $Q^2$
(the EMC effect at large $x$,
nuclear shadowing at small $x$, for a review see e.g. \cite{arneodo}).
Therefore, the separation of nuclear effects from data introduces
certain corrections to QCD analysis of data.

All these motivate us to make a new analysis of the CCFR neutrino data
taking into account corrections due to nuclear effects.
Our analysis involves two steps.
In section \ref{nucl-eff} we discuss our approach to calculate nuclear structure
functions and to correct data for nuclear effects,
and then in section \ref{qcd-fit} we report the results of QCD analysis of corrected data.
In sect. \ref{summary} we summarize.

\section{
Nuclear structure functions}
\label{nucl-eff}

In order to apply corrections for nuclear effects in our analysis we first calculate
the ``EMC ratio'' for the iron target, $R_3(x,Q^2)=F_3^{A}(x,Q^2)/A F_3^N(x,Q^2)$,
with $F_3^A$ the structure function of a heavy nucleus of $A$ nucleons
and $F_3^N$ the structure function of an isolated isoscalar nucleon.%
\footnote{
The isoscalar nucleon structure function is defined as
$F_3^N=\frac12(F_3^p+F_3^n)$.
}
Then we extract the structure function of an isolated isoscalar nucleon
from the CCFR data, $F_3^N(x,Q^2)=F^{\rm CCFR}_3(x,Q^2)/R_3(x,Q^2)$.

Bulk of neutrino data with $Q^2> 1$\,GeV$^2$ is located in the region of
$x>0.1$.
For this kinematical regime it is usually assumed that nuclear DIS of leptons
from nuclear targets can be viewed as incoherent scattering
from bound nucleons.
Major nuclear effects found in this region are due to
nuclear binding \cite{binding}
and Fermi motion \cite{fm}
and off-shell modification of bound nucleon
structure functions \cite{off-shell}.
For the simplest nuclear system, the deuteron, the relation between the deuteron
and the nucleon $F_3$ structure function reads as follows \cite{Ku98},
\begin{eqnarray}
\label{F3D}
xF_3^D(x,Q^2)=2\int\frac{d^3\boldsymbol{p}}{(2\pi)^3}
\left|\Psi_D(\boldsymbol{p})\right|^2\left(1+\frac{p_z}{\gamma M}\right)
x'F_3^N(x',Q^2;p^2),
\end{eqnarray}
where $\Psi_D(\boldsymbol{p})$ is the deuteron wave function which describes
the probability to find the bound proton (or neutron) with momentum $\boldsymbol{p}$,
$x'=Q^2/2p{\cdot}q$ is the Bjorken variable of the bound
nucleon with the four-momentum $p$ which is given by the difference of the target
four-momentum and the four-momentum of the spectator nucleon.
Eq.(\ref{F3D}) is written for the target rest frame
and the axis $z$ is chosen along the direction of momentum transfer,
$q=(q_0,0_\perp,-|\boldsymbol{q}|)$. In this reference frame
$p=(M_D-\sqrt{\boldsymbol{p}^2+M^2},\boldsymbol{p})$
with $M_D$ and $M$ the deuteron and the nucleon mass respectively
and $\gamma=|\boldsymbol{q}|/q_0=(1+4x^2M^2/Q^2)^{1/2}$ is the `velocity'
of the virtual boson.
Note that the bound proton and neutron are off-mass-shell and their structure
functions depend
on the nucleon off-shellness $p^2$ as an additional variable.

For the scattering off a heavy nucleus of $A$ nucleons,
there appears a rich spectrum of spectator nuclear
states of $A{-}1$ nucleons, over which we have to sum.
The nuclear structure function is then given by equation similar to (\ref{F3D})
where we have to substitute the deuteron wave function by nuclear spectral function
${\cal P}(\varepsilon,\boldsymbol{p})$ and introduce an additional integration over
the energy spectrum of spectator states \cite{Ku98},
\begin{eqnarray}
\label{F3A}
xF_3^A(x,Q^2)=\sum_{\tau=p,n}\int\frac{d\varepsilon\,d^3\boldsymbol{p}}{(2\pi)^4}
\,{\cal P}^\tau(\varepsilon,\boldsymbol{p})\left(1+\frac{p_z}{\gamma M}\right)
x'F_3^\tau(x',Q^2;p^2),
\end{eqnarray}
where the sum is over protons ($\tau=p$) and neutrons ($\tau=n$).
The nucleon four-momentum $p=(M+\varepsilon,\boldsymbol{p})$.
The proton and neutron
spectral functions, ${\cal P}^p$ and ${\cal P}^n$, are normalized to the number
of bound protons ($Z$) and neutrons ($N$) respectively.

Heavy nuclei, such as iron $^{56}$Fe$_{26}$, generally have got unequal numbers of protons
and neutrons with an excess of the latter over the former.
The neutron excess is generally small, $(N{-}Z)/A\ll 1$. Therefore, it is a good approximation
to assume that the neutron and the proton spectral functions calculated per one
particle are equal, ${\cal P}^p/Z={\cal P}^n/N$. Then we find from (\ref{F3A}),
\begin{eqnarray}
\label{N-Z}
xF_3^A = \left\langle
\left(1+\frac{p_z}{\gamma M}\right)
\left(x'F_3^N +
\frac{N{-}Z}{2A}\left(x'F_3^n-x'F_3^p\right)
\right)
\right\rangle,
\end{eqnarray}
where
the averaging is done with respect to the isoscalar spectral function,
${\cal P}^p{+}{\cal P}^n$.
The last term in (\ref{N-Z}) gives a correction
due to excess of neutrons in a nucleus. We notice that the sign of this correction
is different for neutrino and anti-neutrino scattering. Indeed,
we have $F_3^\nu=2(d-\bar u)$ with $d$ and $u$
the parton distributions of corresponding quarks in the target (we neglect
for simplicity the contributions due to $s$- and $c$-quarks).
Since the neutron has more $d$-quarks
than the proton has (in the valence quark region),
the neutrino-neutron structure function is larger than the proton one,
$F_3^{\nu n}>F_3^{\nu p}$.
Therefore $F_3^{\nu A}$ receives a positive correction due to excess of
neutrons. Repeating this argument for
anti-neutrino scattering, we find that the corresponding correction
is equal in magnitude but opposite in sign (i.e. negative).
Therefore the $N{-}Z$ correction vanishes for the structure function
averaged over neutrino and antineutrino.
Similar discussion can also be applied to the $F_2$ structure function.
One can also see that the $N{-}Z$ correction is negative for
the charged leptons scattering, i.e. for the $F_2^{\mu A}$ structure function.

As it is obvious from (\ref{F3A}),
calculation of nuclear structure functions requires the knowledge
of nuclear spectral function. In the next section we discuss
the nuclear spectral function used in the present calculation in
more detail.

\subsection{Nuclear spectral function}

Nuclear spectral function ${\cal P}$
determines the probability to find the nucleon with
the momentum ${\boldsymbol p}$ and (non relativistic)
energy $\varepsilon$ in the ground state of the nucleus
and can be written as follows
%
\begin{eqnarray}
\label{SF}
{\cal P}(\varepsilon,{\boldsymbol p})=2\pi \sum_{n,\sigma}\left|
\langle(A-1)_n,{-}{\boldsymbol p}\left|a_\sigma({\boldsymbol p}) \right|A\rangle\right|^2
\delta\left(
\varepsilon+E_n^{A{-}1}+\frac{{\boldsymbol p}^2}{2M_{A{-}1}}-E_0^A\right) .
\end{eqnarray}
Here the sum is over the quantum numbers of the whole set of the residual states
of $A{-}1$ nucleons which includes the bound states as well as the sates in continuum,
$a_\sigma({\boldsymbol p})$ is the annihilation operator of the nucleon with momentum
${\boldsymbol p}$ and polarization $\sigma$, and $E_n^{A{-}1}$ and $E_0^A$ are respectively
the energy of the residual nucleus and the ground state energy of the
target nucleus. The residual system balances momentum of the removed nucleon and
acquires the recoil energy ${\boldsymbol p}^2/2M_{A{-}1}$ though its effect is small for
heavy nuclei.
The nuclear momentum distribution is
\begin{eqnarray}
\label{n}
n({\boldsymbol p})=\int \frac{d\varepsilon}{2\pi}\,
{\cal P}(\varepsilon,{\boldsymbol p}).
\end{eqnarray}
The integration of the spectral function over energy and momentum gives the number
of bound nucleons $A$.

The spectral function (\ref{SF}) determines the rate of nucleon removal
reactions such as $A(e,e'p)X$ that makes it possible to
extract the spectral function from experimental data.\footnote{
Though one should notice, that the direct connection
between the cross sections and the spectral function
holds only in the impulse approximation,
and is destroyed by other effects such as final state interactions
and meson exchange currents.}
The picture of the spectrum of residual states as revealed from these
experiments with heavy nuclei can be summarized as follows.
For small energies and momenta\footnote{
The momenta should be compared with Fermi momentum $p_F$
which is for heavy nuclei $p_F\approx 300$\,MeV/c. The corresponding Fermi
energy is of order $\varepsilon_F\approx 40$\,MeV.}
the energy spectrum of residual states
follows to that predicted by the mean field model of the nucleus, i.e.
it consists of the set of sharp peaks whose positions can be identified with
the energies needed to separate bound nucleons from
the occupied single particle levels in the nuclear mean field.
The deviations from the mean field picture become significant at high momentum ${\boldsymbol p}$
and high nucleon removal energy $\varepsilon$. The widths of the resonances
increase as $\varepsilon$ increases as well as the positions of the peaks
move from that predicted by the energy independent mean field model.
At high energy the spectral function is dominated by contributions from the states
with one and more nucleons in the continuum. These contributions are due to
NN-correlations in the nuclear ground state and can not be accounted for within
the mean field model.

\subsubsection{Phenomenological model of spectral function}

The calculation of the nuclear spectral function for complex nuclei
requires to solve many body problem. The latter is known to be a
difficult task and presently can be done only within certain approximations.
In our discussion we follow \cite{CS96} and consider a phenomenological model for
the spectral function which incorporates both the single particle
nature of the spectrum at low energy as well as high energy and high
momentum components due to NN-correlations in the ground state.%
\footnote{
We note that our definition of the spectral function is different from the one
used in \protect\cite{CS96}, where the recoil energy was not included into
the energy $\delta$-function in (\protect\ref{SF}).
}
To this end we separate the full spectral function (\ref{SF}) into two parts,
\begin{eqnarray}
\label{01}
{\cal P}(\varepsilon,{\boldsymbol p})={\cal P}_0(\varepsilon,{\boldsymbol p})
+{\cal P}_1(\varepsilon,{\boldsymbol p}),
\end{eqnarray}
which correspond to contributions from low excitation energy intermediate states
(${\cal P}_0$) and high excitation energy states (${\cal P}_1$).
The low energy part can be approximated by the sum of the energy $\delta$-functions
which pick the positions of the occupied single particle levels
weighted with the corresponding wave functions squared.
In practice we use an approximate expression instead where the sum
over occupied levels is substituted by its average value,
\begin{eqnarray}
\label{P0}
{\cal P}_0(\varepsilon,{\boldsymbol p})=2\pi\, n_0({\boldsymbol p})
\delta\left(\varepsilon+E^{(1)}+\frac{{\boldsymbol p}^2}{2M_{A{-}1}}\right),
\end{eqnarray}
with $E^{(1)}=E^{A{-}1}-E_0^A$ the nucleon separation energy
averaged over residual configurations of $A{-}1$ nucleons
with low excitation energies, i.e. mean field configurations,
and $n_0({\boldsymbol p})$ the corresponding part of the nucleon
momentum distribution.

The high energy part ${\cal P}_1$ is determined by excited states
in (\ref{SF}) with one or more nucleons in the continuum.
It was observed within many body calculations \cite{md,CS96}
for a wide range of nuclei
that nuclear momentum distributions at high momenta
$(|{\boldsymbol p}|>p_F$ with $p_F$ the Fermi momentum)
run parallel to
the deuteron distribution $n_D({\boldsymbol p})$,
\begin{eqnarray}
\label{n1-nD}
n_1({\boldsymbol p})\approx C^A n_D({\boldsymbol p}),
\end{eqnarray}
where the normalization
constant $C^A$ incorporates the many body aspects of the problem.
It was found \cite{CS96}
that the constants $C^A$ increase from $2$ for $^3$He
to $4.5$ for $^{56}$Fe. Going to a larger mass number does not bring in more
high momentum component, $C^A=5$ for nuclear matter.%
\footnote{
One should note however, that the relation (\ref{n1-nD}) does not hold
at low momentum where $n_1({\boldsymbol p})$ contributes only a little
to the full momentum distribution.  }

This observation finds a simple interpretation if one assumes that
high momentum component is generated by ground state configurations
with a correlated NN-pair with a small distance between the nucleons.
One can expect therefore that the relative motion in the NN-pair is
determined by the properties of the NN-interaction in the vacuum
rather than by long range nuclear interactions, and the distribution
in the relative momentum will be similar to momentum distribution in
the deuteron.

In terms of the spectral function ${\cal P}_1$ this corresponds to
the assumption about the dominance of the contribution from the states
with one nucleon in the continuum
and the remaining $A{-}2$ nucleons being in a state with
low momentum and low excitation energy,
\begin{eqnarray}
\label{final-st}
|A{-}1,-{\boldsymbol p}\rangle \approx
a^\dagger({\boldsymbol p}_1)|(A{-}2)^*,{\boldsymbol p}_2\rangle
\delta({\boldsymbol p}_1+{\boldsymbol p}_2+{\boldsymbol p}) .
\end{eqnarray}
The coprresponding matrix element in (\ref{SF}) is then determined by the wave
function of the NN-pair embeded into nuclear environment,
\begin{eqnarray}
\label{factor}
\left\langle (A{-}2)^*,{\boldsymbol p}_2\left|
a({\boldsymbol p}_1)a({\boldsymbol p})
\right|A\right\rangle =
\psi_{\rm rel}({\boldsymbol k})
\psi_{\rm CM}^{A{-}2}({\boldsymbol p}_{\rm CM})
\delta({\boldsymbol p}_1+{\boldsymbol p}_2+{\boldsymbol p}).
\end{eqnarray}
We assume here factorization into the wave functions describing
the relative motion in the NN pair, $\psi_{\rm rel}({\boldsymbol k})$,
with relative momentum
${\boldsymbol k}=({\boldsymbol p}-{\boldsymbol p}_1)/2$
and the center-of-mass
(CM) motion of the pair in the field of $A{-}2$ nucleons,
$\psi_{\rm CM}^{A{-}2}({\boldsymbol p}_{\rm CM})$ with
${\boldsymbol p}_{\rm CM}={\boldsymbol p}_1+{\boldsymbol p}$.
In general $\psi_{\rm CM}$ depends on the quantum
numbers of the state of $A{-}2$ nucleons, however the
corresponding dependence of the $\psi_{\rm rel}$ is weak.

We substitute (\ref{factor}) into (\ref{SF})
and sum over the spectrum of $A{-}2$ nucleons states and
obtain an approximate expression for ${\cal P}_1$,
\begin{eqnarray}
\label{P1}
{\cal P}_1(\varepsilon,{\boldsymbol p})&=&(2\pi)
\int d^3{\boldsymbol p}_1 d^3{\boldsymbol p}_{\rm CM}
n_{\rm rel}({\boldsymbol k}) n_{\rm CM}({\boldsymbol p}_{\rm CM})
\delta({\boldsymbol p}_1+{\boldsymbol p}-{\boldsymbol p}_{\rm CM}) \\ \nonumber
&& \delta\left(\varepsilon+\frac{{\boldsymbol p}_1^2}{2M}+
	\frac{{\boldsymbol p}_{\rm CM}^2}{2M_{A{-}2}}+E^{(2)}\right).
\end{eqnarray}
Here $n_{\rm rel}$ and $n_{\rm CM}$ are the relative and the CM momentum
distributions respectively and $E^{(2)}=E^{A{-}2}-E_0^A$ is the energy
needed to separate two nucleons from the ground state averaged over
configurations of $A{-}2$ nucleons with low excitation energy.
Note that the minimum two nucleon  separation energy
$E^{(2)}=E_0^{A{-}2}-E_0^A$
is of order of $20\,$MeV for medium range nuclei like $^{56}$Fe.

The factorization of the matrix element (\ref{factor}) into the relative
and the CM motion wave functions is justified if relative momentum in
the NN-pair is large relative to the CM momentum of the pair. This can be written
as $|{\boldsymbol p}|\gg |{\boldsymbol p}_{\rm CM}|$.
This condition allows us to approximate (\ref{P1})
by taking the relative momentum distribution out of the integral over
the CM momentum at the point ${\boldsymbol k}={\boldsymbol p}$.  Then
we have,
\begin{eqnarray}
\label{P1-av}
{\cal P}_1(\varepsilon,{\boldsymbol p})&=&(2\pi)n_{\rm rel}({\boldsymbol p})
\left\langle
\delta\left(\varepsilon+\frac{({\boldsymbol p}+{\boldsymbol p}_2)^2}{2M}+
	\frac{{\boldsymbol p}_2^2}{2M_{A{-}2}}+E^{(2)}\right)
\right\rangle_{\rm CM} .
\end{eqnarray}
where the averaging is done with respect to the CM motion of the pair.
From the latter equation it is clear that the high momentum part of
nuclear momentum distribution is given by the relative momentum
distribution in the correlated NN pair embedded into nuclear
environment, $n_1({\boldsymbol p})=n_{\rm rel}({\boldsymbol p})$.

The characteristic momentum for the CM motion of the NN-pair is
similar to the one in the mean field model. In fact the averaged CM
momentum squared of the pair  can be estimated from the balance of the
overall nucleus momentum \cite{CS96},
$\langle(\sum {\boldsymbol p}_i)^2\rangle=0$,
where the sum is taken over all bound nucleons and
the averaging is performed with respect to the intrinsic wave
function of the nucleus. This gives $\langle{\boldsymbol p}^2_{\rm
CM}\rangle= 2(A{-}2)\langle{\boldsymbol p}^2\rangle/(A{-}1)$, with
$\langle{\boldsymbol p}^2\rangle$ the mean value of the squared single
nucleon momentum.
Since, by our assumption, the CM distribution does not include
high-momentum component, we should also exclude the contribution of high-momentum part
in estimating $\langle{\boldsymbol p}^2\rangle$.
We follow \cite{CS96} and parameterize the CM momentum distribution of
the correlated NN pair in the field of other $A{-}2$ nucleons by a
Gaussian distribution,
\begin{eqnarray}
\label{nCM}
n_{\rm CM}({\boldsymbol p}_{\mathrm CM})=\left({\alpha}/{\pi}\right)^{3/2}
\exp(-\alpha {\boldsymbol p}_{\mathrm CM}^2),
\end{eqnarray}
with the parameter $\alpha$ determined from the averaged CM momentum of the pair,
$\alpha=3/(2\langle{\boldsymbol p}^2_{\rm CM}\rangle)$.

Using (\ref{nCM}) we find that the integration over the CM momentum in (\ref{P1-av})
can be done analytically and finally the result reads,
\begin{eqnarray}
\label{P1-gauss}
{\cal P}_1(\varepsilon,{\boldsymbol p})=n_1({\boldsymbol p})\frac{2M}{|{\boldsymbol p}|}
\sqrt{\alpha\pi}
\left(\exp({-\alpha p_{\rm min}^2})-\exp({-\alpha p_{\rm max}^2})\right),
\end{eqnarray}
where $p_{\rm min}$ and $p_{\rm min}$
are respectively the minimum and the maximum CM momenta allowed by the energy-momentum
conservation in (\ref{P1}) for the given $\varepsilon$ and ${\boldsymbol p}$,
\begin{eqnarray}
&&
p_{\rm max}^2=\left(\frac{A{-}2}{A{-}1}|{\boldsymbol p}|+ p_T\right)^2,\quad
p_{\rm min}^2=\left(\frac{A{-}2}{A{-}1}|{\boldsymbol p}|- p_T\right)^2,
\\
\nonumber
&&
\mbox{with}\quad
{p_T}=\left(
\frac{A{-}2}{A{-}1}\left(
-2M(\varepsilon + E^{(2)})-\frac{{\boldsymbol p}^2}{{A{-}1}}
\right)\right)^{1/2}.
\end{eqnarray}
We notice that $p_T$ has the interpretation of the maximal allowed CM momentum
in the correlated NN-pair in the direction transverse to ${\boldsymbol p}$
for the fixed $\varepsilon$ and $|{\boldsymbol p}|$.
Note that the separation energy $\varepsilon$ is negative, as it follows from its definition
in (\ref{SF}). The condition $p_T^2=0$ determines the threshold value of $\varepsilon$
for the fixed $|{\boldsymbol p}|$.


In numerical evaluations we use the parameterizations for $n_0({\boldsymbol p})$ and
$n_1({\boldsymbol p})$
of \cite{CS96}, which fit nicely the results of many body calculation of nuclear momentum
distribution. It follows from this calculation that
low momentum part incorporates about 80\% of the total normalization
of the spectral function while the other 20\% are taken by the high momentum part.
The mean kinetic energy obtained from integration of the full momentum distribution
$n_0+n_1$ for the iron nucleus
is $\langle{\boldsymbol p}^2\rangle/2M=31\,$MeV (the share of the high-momentum
component $n_1$ is about 20\,MeV).
The two parameters, $E^{(1)}$ and $E^{(2)}$, determine characteristic range of
nucleon separation energy. We set $E^{(2)}=E_0^{A{-}2}-E_0^A=20\,$MeV, and therefore
neglect possible contributions due to excited states of $A{-}2$ nucleons in (\ref{P1}).%
\footnote{
The effect of $A{-}2$ excited states would lead to an overall increase
of nucleon separation energy. We believe, however, that concrete estimates
of this effect would require us to go beyond the model discussed in the present
paper}
In order to fix the parameter $E^{(1)}$ we employ the Koltun sum rule \cite{KSR},
which gives the relation between  mean separation energy $\langle\varepsilon\rangle$,
mean kinetic
energy $\langle{\boldsymbol p}^2\rangle/2M$,
and the ground state energy per nucleon
$E_0^A/A$.
For the mean kinetic energy of $31\,$MeV the sum rule gives
$\langle\varepsilon\rangle\approx -50\,$MeV. By integrating our model spectral
function we find the value $E^{(1)}=27\,$MeV which satisfies the Koltun sum rule.

\subsection{The EMC ratios $R_2$ and $R_3$}

\begin{figure}[p]
\begin{center}
%
\includegraphics[width=\textwidth]{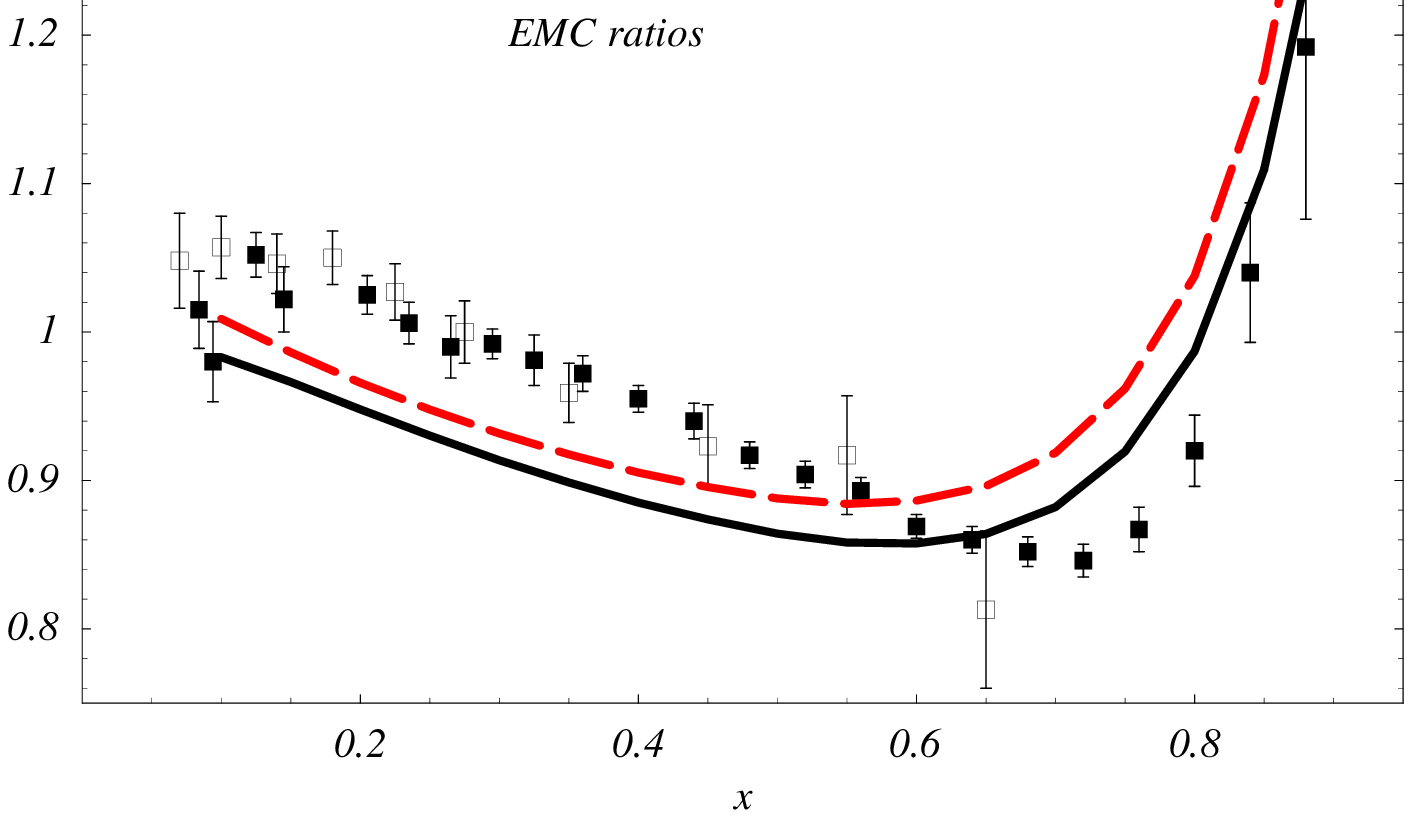}
\caption{
The iron/deuterium ratios (EMC ratios)
calculated for the structure functions $F_2$ (solid curve)
and $F_3$ (dashed curve) within the model described in the text.
The data points are from BCDMS \protect\cite{bcdms}
(open boxes) and SLAC \protect\cite{slac} (filled boxes) experiments.
The curves were calculated at fixed $Q^2=16\,$GeV$^2$ using CTEQ4 parameterizations
for the nucleon parton distributions.
\label{emc_ratios}
}
\end{center}
\end{figure}

In Fig.\ref{emc_ratios} we compare the iron/deuterium ratios for the charged lepton
structure
function $F_2^\mu$ and the neutrino and antineutrino averaged
structure function $F_3$, calculated with the model
nuclear spectral function discussed above.
Also shown are the BCDMS \cite{bcdms} and the SLAC \cite{slac}
data on the iron/deuterium
$F_2$ structure function ratios. In numerical calculations we
use the CTEQ4 parameterizations for the nucleon parton
distributions \cite{cteq}.
We see that the behavior of the ratio $R_3$ is very similar
to that of the ratio $R_2$ (for large $x$ and large $Q^2$
this does not come as a big surprise, since both $F_2$ and $xF_3$ are determined
by valence quarks in this region). A small difference between the $R_2$ and
$R_3$ curves is due to the neutron excess correction. As it was discussed in sect. \ref{intro},
the $F_2^\mu$ structure function receives a negative $N{-}Z$ correction, while
similar correction cancels out in the neutrino and antineutrino averaged structure functions.
It is well known, that the depletion of nuclear structure functions at $x<0.7$
is due to nuclear binding effect \cite{binding}, while
the rise of the ratios at large $x>0.7$ is due to nuclear momentum distribution
effect (Fermi motion).

We recall also that bound nucleons are off-mass-shell.
Off-shell effects in the structure functions appear as the dependence
on the target invariant mass $p^2$. Target mass corrections can be of two
different kinds. First of all, we have to take into account `kinematical'
target mass dependence due to finite $p^2/Q^2$ ratio.
To this end we use the Nachtmann
scaling variable \cite{Nacht}
$\xi=2x'/(1+(1+4x'^2p^2/Q^2))^{1/2}$
instead of the Bjorken variable $x'$.
Other (`dynamical') sources of $p^2$-dependence of structure functions
are also possible. In this respect we refer to a model where $p^2$-dependence of
structure functions appears in the leading order \cite{off-shell,Ku98}.
We note also here, that we take into account off-shell effects in
the bound nucleon structure function in a way that it does not
affect the number of valence quarks in the nucleon \cite{Ku98}.
The off-shell effect
acts coherently with
nuclear binding effect and leads to an additional suppression of
nuclear structure functions at intermediate range of $x$.

The ratio $R_2$ follows quite closely to data on the EMC effect in the iron
nucleus (see Fig. \ref{emc_ratios}). This gives us the confidence in our method
to calculate the EMC effect in the $F_3$ structure function.

\section{QCD analysis and fit}
\label{qcd-fit}

Our QCD fit proceeds as follows.
The nucleon structure function  $F^N_3(x,Q^2)$ is written as a sum of the leading twist
and the high twist terms, (\ref{LTplusHT}).
 We parametrize $xF_{3}^{LT}$
at some  scale $Q^2=Q_0^2$
in terms of a simple function,
\begin{eqnarray}
xF_3^{LT}(x,Q^2_0) =a_1x^{a_2}(1-x)^{a_3}(1+a_4 x).
\label{xf30}
\end{eqnarray}
Then we apply the renormalization group equation in order to calculate evolution of
$xF_{3}^{LT}$ with $Q^2$. We solve the renormalization group equation in the
leading (LO), next-to-leading (NLO) and next-to-next-to-leadig (NNLO) logarithm
approximations of QCD. In doing so we expand the leading twist
structure function $xF_3^{LT}$ in terms of its
Mellin moments within the framework of the Jacobi polinomial method and then apply
the evolution equations to the moments (for more detail on the method used see
\cite{jacobi-pol}).

It should be noticed that in general the higher twist terms can be of two kinds:
those which have the kinematical
nature, e.g. the terms due to finite target mass, and those which arise due to
higher twist operators and reflect the quark-gluon interaction effects in the
target (``pure'' higher twists).
In order to ensure that the higher twist term in (\ref{LTplusHT})
describes effects due to quark-gluon
interaction in the target
we explicitly take into account
the kinematical corrections due to finite target mass.
To this end we substitute
the Mellin moments by the Nachtmann moments
\cite{Nacht} in the the Jacobi polinomial expansion
of the leading twist structure function $xF_{3}^{LT}$.

The CCFR data points are given in terms of discrete $x$-bins structure which range
from $x=0.0075$ to $x=0.75$.
We fit 116 data points with $Q^2$ in the range between
$1.3\,$GeV$^2$ and $200\,$GeV$^2$. The fit parameters are the parameters
$a_2$, $a_3$, and $a_4$ of
(\ref{xf30}) at the scale $Q^2_0$, the values of the function $h(x_i)$ at the center of
each $x_i$-bin, as well as the QCD scale parameter  $\Lambda_{\overline{MS}}$.
We fix the parameter $a_1$ by normalizing (\ref{xf30}) to the
Gross-Llewellyn-Smith sum rule, which was calculated in QCD to the second order
 in $\alpha_S$ \cite{GL},
$S_{\rm GLS}= 3(1-\alpha_S/\pi-3.25(\alpha_S/\pi)^2)$.

\begin{figure}[p]
\begin{center}
\vspace{-2cm}
 \includegraphics[width=0.85\textwidth]{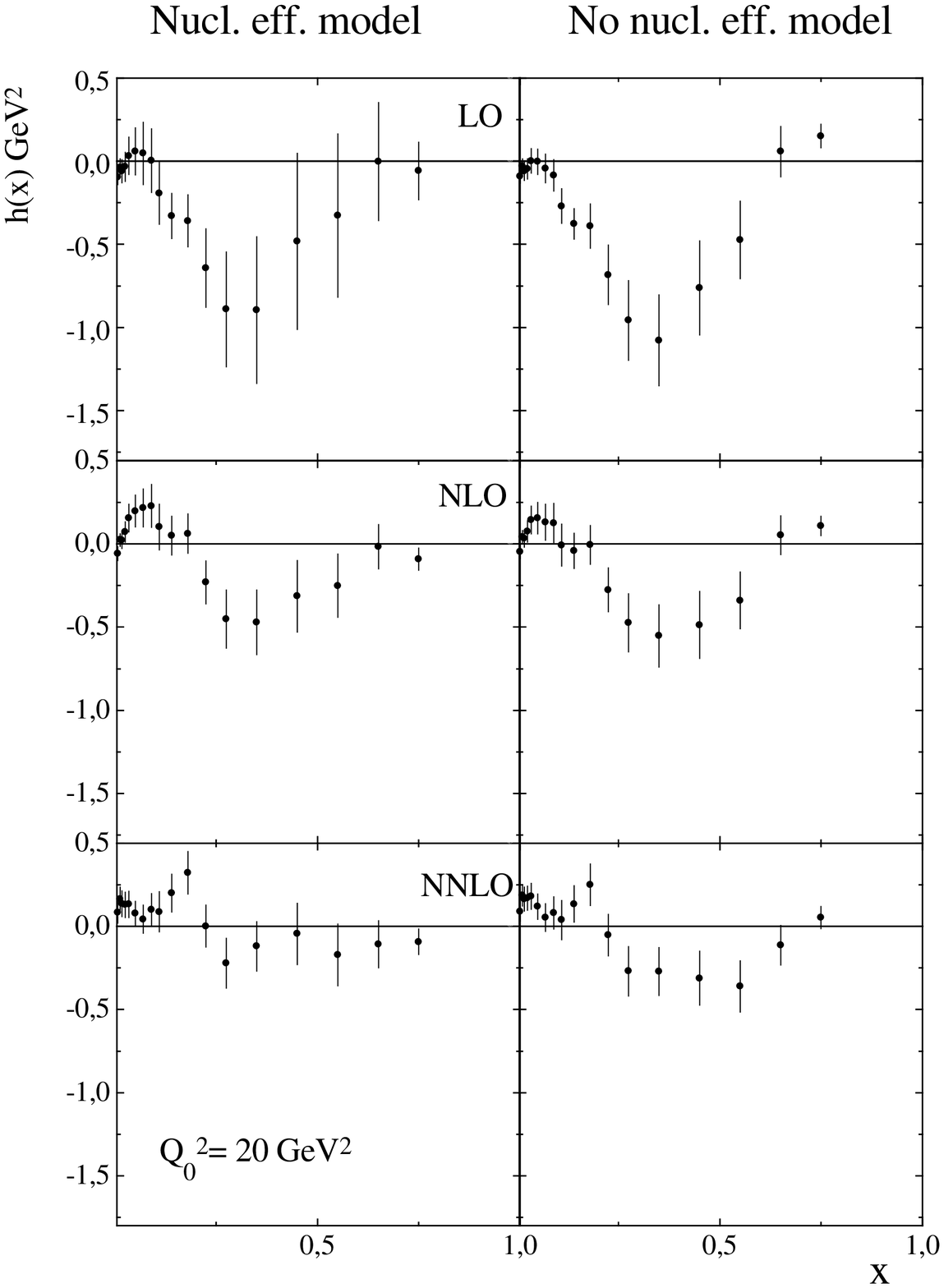}
%
\caption{
The function $h(x)$, which describes the strength of the higher twist
term in the $xF_3$ structure function as extracted from the fit to the
CCFR neutrino data (see text).
The labels on the figure indicate the level to which the perturbation theory
analysis of $xF_3^{LT}$ is done.
\label{ht}
}
\end{center}
\end{figure}

Our results are shown in Fig. \ref{ht} for the LO, NLO and NNLO
approximations to the evolution equation. Also shown are the results
with and without applying corrections for nuclear effects.  We found
that the fitting parameters are stable for $Q_0^2>15\,$GeV$^2$ and
have chosen $Q_0^2=20\,$GeV$^2$ for the results presented in
Fig. \ref{ht}.
The present fit includes more experimental points than that of
\cite{fit1,fit2}. In particular, the inclusion of low-$Q^2$ data points into the fit
allows us to reduce the error bars in $h(x)$ as compared to those presented in
\cite{fit1,fit2}.
Though we should notice some increasing theoretical uncertainties associated
with low $Q^2$ data included into our analysis.
We found that, in general, our present fit agrees with \cite{fit1,fit2},
though introduces certain corrections especially at large $x$.

A special care was taken to insure that our method to separate target mass
correction is self-consistent. In particular we have done a special fit with
$h(x_i)$ fixed at values presented in Fig. \ref{ht}, but let the mass parameter
in the Nachtmann moments to be free. We found that the minimum of $\chi^2$
corresponds to the value of the mass parameter about $0.9\,$GeV with an error
about $0.2\, $GeV. This value only weakly depends on the order of perturbative
analysis and is close to the proton mass, that gives us confidence in the method
used.


As one can clearly see from Fig. \ref{ht} the magnitude of the higher twist effects
depends on the level to which
the perturbation theory analysis of $F_3^{LT}$ is done.
The more perturbative corrections are included into the evolution equation,
the less room is left for the function $h(x)$.
The shape of $h(x)$ in NLO is in a qualitative agreement with
 the prediction of
infrared renormalon approach \cite{h-function}.
In particular, we found that the function $h(x)$ is negative
in the region $0.1<x<0.6$.

The separation of nuclear effects from  data leads to a further suppression of
the higher twist term $h(x)$ at all levels of perturbation theory analysis of
$F_3^{LT}$.
The effect of nuclear corrections on $h(x)$ is most pronounced at large $x$, where we
observe a systematic reduction of $h(x)$ as compared with no-nuclear-effects analysis.
Nuclear corrections result in the decrease of the values of the function $h(x)$ at
$x>0.6$. As one can see from Fig. \ref{ht}, the central points of
$h(x)$ at $x=0.65$ and $x=0.75$ bins become negative
in contrast to the infrared renormalon prediction for large $x$ \cite{h-function}.
An attempt to take into account nuclear effects in the QCD fit was previously done in
\cite{ST97,fit2}.
We comment in this respect that authors of
\cite{ST97} used the deuteron model for nuclear effects,
which is not a realistic one for the iron target.
In particular, we found that $h(x)$ is negative for large $x$, while it was  positive in
\cite{ST97}.
Authors of \cite{fit2} attempted to introduce nuclear corrections to QCD fit in terms
of the moments of structure functions. However, it was incorrectly
assumed in \cite{fit2}, that the nuclear structure function
$F_3^A\to 0$ as $x\to 1$,
that, to our mind, caused the $\chi^2$ increase in their QCD fit.

We found that the scale parameter $\Lambda_{\overline{MS}}$ decreases
for about $40\,$MeV after nuclear effects are taken into account.
This will lead to a shift of $\alpha_S(M_Z)$ for about $2\cdot 10^{-3}$.
Within the NNLO fit we get $\Lambda_{\overline{MS}}=(394 \pm
55)\,$MeV for four quark flavors.

\section{Summary and conclusions}
\label{summary}

In the present paper we report the results of our QCD analysis of
CCFR data on $F_3$ structure function. The main emphasis was put
on the extraction of higher twist contribution from data.
We took special care to separate nucler effects from data,
and compare the results of both analyses with and without
corrections for nuclear effects.

We found that nuclear effects cause about 10\% decrease in the
$\Lambda_{\overline{MS}}$ value.

Our analysis confirms the observation made earlier, that the
magnitude of higher twist terms decreases
strongly when going from LO to NLO, and then to NNLO,
approximations to the evolution equation.
We observe an additional suppression of higher twist terms
when corrections due to nuclear effects have been applied.

In conclusion we note that small-$x$ region in $F_3$ structure
function is of particular interest, where a strong nuclear shadowing effect
is anticipated \cite{shadow}.
We plan to address nuclear shadowing effect in application to QCD analysis
of neutrino data.

\section{Acknowledgements}

This work was supported in part by the RFBR project no. 00-02-17432.
We are grateful to A. L. Kataev for useful discussions.



\end{document}